\documentclass[%
 aip,
rsi,%
amsmath,amssymb,
preprint
]{revtex4-1}

\usepackage{graphicx}
\usepackage{dcolumn}
\usepackage{bm}

\begin{document}

\title[Interband transitions in narrow-gap carbon nanotubes and graphene nanoribbons]{Interband transitions in narrow-gap carbon nanotubes and graphene nanoribbons}

\author{R. R. Hartmann}
\affiliation{
Physics Department, De La Salle University,  2401 Taft Avenue,
0922 Manila, Philippines
}

\author{V. A. Saroka}
\affiliation{Institute for Nuclear Problems, Belarusian State University, Bobruiskaya 11, 220030 Minsk, Belarus
}

\author{M. E. Portnoi}
\email{M.E.Portnoi@exeter.ac.uk}
\affiliation{Physics and Astronomy, University of Exeter, Stocker Road, Exeter EX4 4QL, United Kingdom}
\affiliation{ITMO University, St. Petersburg 197101, Russia}

\date{25 March 2019}

\begin{abstract}
We use the robust nearest-neighbour tight-binding approximation to study on the same footing interband dipole transitions in narrow-bandgap carbon nanotubes and graphene nanoribbons. It is demonstrated that curvature effects in metallic single-walled carbon nanotubes and edge effects in gapless graphene nanoribbons not only open up bang gaps, which typically correspond to THz frequencies, but also result in a giant enhancement of the probability of optical transitions across these gaps. Moreover, the matrix element of the velocity operator for these transitions has a universal value (equal to the Fermi velocity in graphene) when the photon energy coincides with the band-gap energy. 
Upon increasing the excitation energy, the transition matrix element first rapidly decreases (for photon energies remaining in the THz range but exceeding two band gap energies it is reduced by three orders of magnitude), and thereafter it starts to increase proportionally to the photon frequency. A similar effect occurs in an armchair carbon nanotube with a band gap opened and controlled by a magnetic field applied along the nanotube axis. There is a direct correspondence between armchair graphene nanoribbons and single-walled zigzag carbon nanotubes. The described sharp photon-energy dependence of the transition matrix element together with the van Hove singularity at the band gap edge of the considered quasi-one-dimensional systems make them promising candidates for active elements of coherent THz radiation emitters. 
The effect of Pauli blocking of low-energy interband transitions caused by residual doping can be suppressed by creating a population inversion using high-frequency (optical) excitation.
\end{abstract}

\keywords{carbon nanotubes; graphene nanoribbons; interband transitions; THz radiation}
\maketitle

\section{Introduction}\label{sec:Intro}
Creating reliable and portable coherent sources and sensitive detectors of terahertz (THz) radiation is one of the most formidable tasks of modern device physics.~\cite{Lee2007}  Potential applications of THz spectroscopy range from medical imaging and security to astrophysics and cosmology. The unique position of the THz range, in the gap between the parts of electromagnetic spectrum, which are accessible by either electronic or optical devices, leads to an unprecedented diversity in approaches to bridging the gap.\cite{Lee2007,Ferguson2002,Dragoman2004,Mangeney2007,Avrutin1988,Kruglyak2005,Mikhailov2007a,Nagatsuma2011,Mangeney2014,Kusmartsev2017} One of the latest trends in THz technology is to employ carbon nanomaterials as building blocks of high-frequency devices.\cite{Hartmann2014}  In particular, there are a growing number of proposals using carbon nanotubes for THz applications including several schemes\cite{Kibis2005,Kibis2005b,Kibis2007,Portnoi2008,PORTNOI2009,Rosenau2009} put forward by some of the authors of the present work.

Within the frame of a simple zone-folding model of the $\pi$-electron graphene spectrum, all single wall carbon nanotubes (CNTs) that satisfy the condition $n=3p+m$, where $p$ is an integer, are metallic (see Refs.~\cite{SaitoBook1998,Dragoman2004} for classification of CNTs). However, first principle and numerical tight-binding calculations show that only armchair CNTs ($n=m$) are truly metallic.~\cite{Mintmire1992,Reich2002a,Lo2010} All other tubes from the specified category have a small curvature-induced band gap that ranges from $\approx2-50$ meV depending on the tube diameter and chirality. Thus, CNTs characterized by the indices $n=3p+m$ are commonly referred to as quasi-metallic CNTs, and the presence of the band gap in quasi-metallic CNTs has been detected by scanning tunneling microscopy and electrical transport measurements.\cite{Ouyang2001,Zhou2000}

Many THz/far-infrared spectroscopy experiments have been performed on CNTs.~\cite{Bommeli1996,Ugawa1999,Itkis2002,Jeon2005,Jeon2004,Jeon2002,Akima2006,Borondics2006,Nishimura2007,Ren2009a,Slepyan2010,Shuba2012,Ren2012a,Ren2013,Zhang2013} Several groups observed a broad terahertz absorption peak, the origin of which was attributed to interband absorption in quasi-metallic CNTs with curvature-induced gaps.~\cite{Ugawa1999,Borondics2006,Nishimura2007,Kampfrath2008} However, an alternative explanation where absorption was attributed to collective electron excitations such as plasmon resonances\cite{Jeon2002,Slepyan2010,Shuba2012,Akima2006,Nakanishi2009,Ren2013,Zhang2013} was also put forward. With over a decade of scientific argument and many reports of controversial and contradictory results from different groups, it seems that the prevailing consensus is that at high carrier densities the THz peak is due to collective intraband effects rather than single electron interband optical transitions.

Despite the enormous attention curvature effects have received in relation to the electronic band structure of nanotubes, the role of curvature in regards to their optical properties has garnered considerably less attention, with notable exceptions including proposals for THz radiation emitters.~\cite{Kibis2007,PORTNOI2009} Interband transitions in quasi-metallic tubes in the THz regime are allowed even in the absence of curvature effects. However, these transitions are very weak at low frequencies, owing to the fact that the matrix element of velocity is proportional to $a\nu$,\cite{Kibis2007} where $a$ is the graphene lattice constant and $\nu$ the frequency of excitation. In what follows, it is shown that the same curvature effect in quasi-metallic CNTs which opens the gap in the nanotube energy spectrum also allows strong interband transitions in the THz range. These transitions are several orders of magnitude larger then those previously considered in a model, which neglected curvature.\cite{Kibis2007} This is because the inclusion of curvature effects results in the matrix element of velocity becoming equal to $v_{\mathrm{F}}$; which means the optical transitions in the vicinity of the Dirac point are as strong as allowed optical transitions between more distant subbands. We also show that the Aharonov-Bohm effect reported previously for tubes without taking into account curvature\cite{Ajiki1993,Ajiki1996} results in the splitting of the THz peak associated with curvature. Furthermore, controlling the strength of the magnetic field directed along the nanotube axis allows the position of the THz peaks to be tuned.

Graphene nanoribbons (GNRs) represent another type of quasi-one-dimensional carbon nanostructures and can be imagined as narrow stripes cut from a single layer graphene sheet. Just as how the rolling of the tube determines its optical and electric properties, the manner in which the ribbon is cut is equally as important. The highest symmetry nanoribbons are formed by ``cutting" along parallel lines to form either zigzag or armchair edges, whence the origin of their names. These ribbons are specified by the number of carbon atoms pairs $N$, or equivalently by the number of ``zigzag lines" for zigzag or ``dimer lines" for  armchair nanoribbons. The most simple tight-binding model shows that all zigzag ribbons (ZGNR) are metallic, whereas only armchair ribbons (AGNRs) with $N=3p+2$, where $p$ is an integer, are gapless. The low-energy dispersion of electrons in metallic AGNRs is linear and similar to that of metallic CNTs, while the electron dispersion of ZGNRs is dominated by edge states.~\cite{fujita1996peculiar,PhysRevB.54.17954,PhysRevB.59.8271} However, in actuality, both types of the metallic ribbons are quasimetallic. The electron dispersion of ZGNR edge states is strongly modified by electron-electron interaction, whereas for AGNR the energy dispersion is influenced by the change of C-C bonds at the edge of the ribbon compared to bonds in the ribbon interior. In both cases the outcome is a small band gap opening of the order of $50$ meV. In what follows we consider only quasimetallic AGNRs, for which the edge effects lead to more prominent interband transitions between the closest valence and conduction subbands than for ZGNRs.\cite{Saroka2017} 
As in the case of quasi-metallic CNTs, the interband matrix element of velocity of narrow-gap AGNRs is equal to the Fermi velocity at the band gap edge. This, coupled with the van-Hove singularity in the joint density of states makes both quasi-metallic CNTs and AGNRs promising candidates as the building block of high-frequency devices.

\section{Carbon Nanotubes}\label{sec:CNTs}
\subsection{The band structure of narrow-gap CNTs with curvature}

The rolling of a graphene sheet to form a carbon nanotube has three main consequences:\cite{Charlier2007} C-C bond length contraction, the rotation of the $2p_{z}$ orbitals and the rehybridization of the $\pi$ and $\sigma$ orbitals. All of the aforementioned effects result in the modification of the hopping parameters of the tight-binding Hamiltonian.\cite{Kane1997,Kleiner2000,Kleiner2001} In the present study we focus on the effect of bond length contraction. 

The tight-binding Hamiltonian of a graphene-like 2D crystal can be written as:
\begin{equation}
H \left( \boldsymbol{k} \right) = \left(\begin{array}{cc}
0
& f\left(\boldsymbol{k}\right)\\
f^{\ast}\left(\boldsymbol{k}\right) 
& 
0
\end{array}\right) \, ,
\label{eq:Honeycomb_HAM}
\end{equation}
where $f\left(\boldsymbol{k}\right)=\sum_i t_i \exp\left(i \boldsymbol{k} \cdot \boldsymbol{R}_i\right)$, $\boldsymbol{k}$ is the charge carrier's wavevector, $\boldsymbol{R}_i$ are the nearest neighbor vectors, and $t_i$ are their associated hopping integrals,~\cite{SaitoBook1998} which for graphene are equivalent i.e., $t_{i}=t\approx-3$ eV. For a pristine graphene sheet the nearest neighbor vectors are defined as $\boldsymbol{R}_1=\left(\frac{a}{\sqrt{3}},0\right)$, $\boldsymbol{R}_2=\left(-\frac{a}{2\sqrt{3}},-\frac{a}{2}\right)$ and $\boldsymbol{R}_3=\left(-\frac{a}{2\sqrt{3}},\frac{a}{2}\right),$ where $a=\sqrt{3}a_{\mathrm{CC}}$ and $a_{\mathrm{CC}}$ is the nearest neighbor distance between two carbon atoms which is given as $1.42\,\textrm{\AA}$. The Hamiltonian, Eq.~(\ref{eq:Honeycomb_HAM}), acts on the basis $\left(\left|\psi_{A}\right\rangle ,\,\left|\psi_{B}\right\rangle \right)^{T}$, where $\left|\psi_{A}\right\rangle$ and $\left|\psi_{B}\right\rangle$ are the tight-binding wavefunctions associated with the two sub-lattices, to yield the eigenvalues $\xi_{j} = s|f| $,
where $\tilde{s}=+1$ for the conduction band and $\tilde{s}=-1$ for the valence band.
\begin{figure}
\centering
\includegraphics*[width=0.8\linewidth]{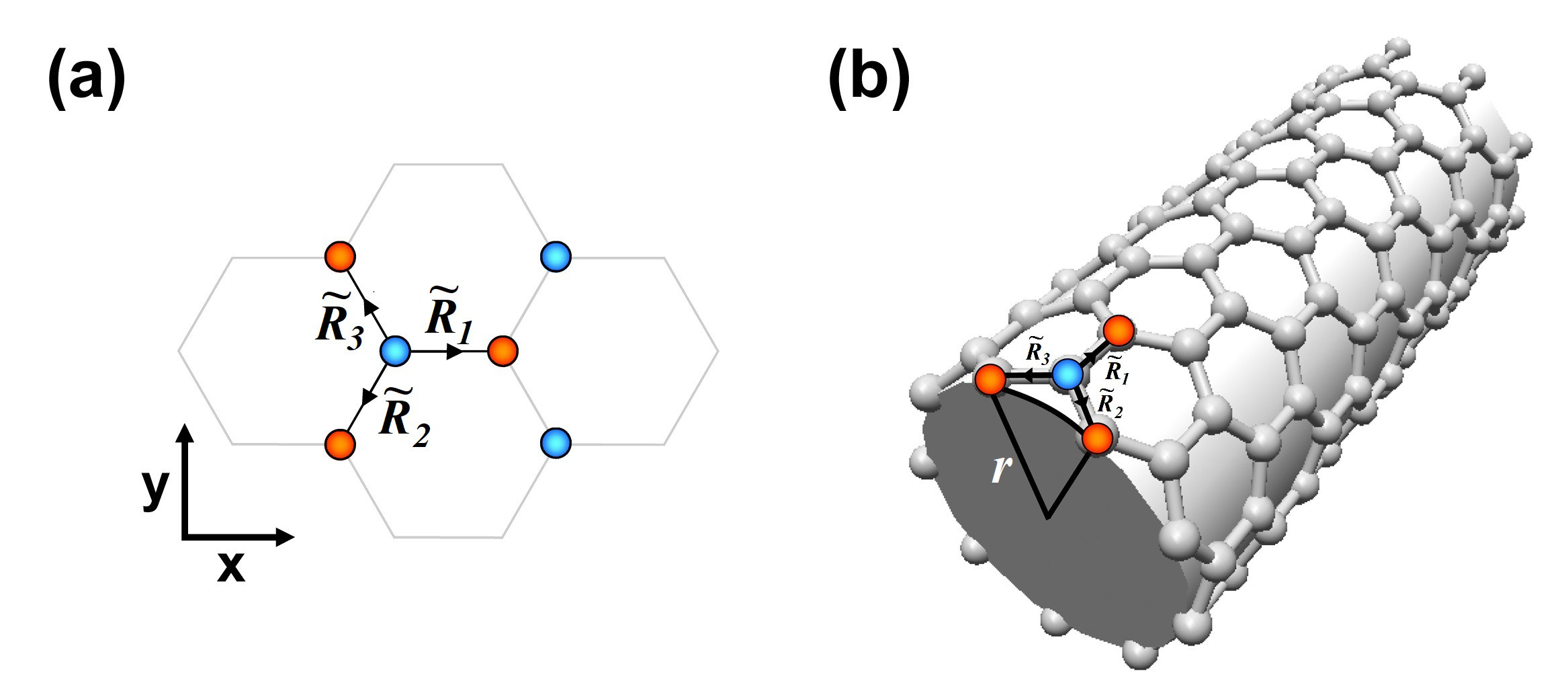}
\caption{
(a) A carbon nanotube as an unrolled graphene sheet, represented by the set of effective nearest neighbor vectors $\widetilde{\boldsymbol{R}}_i$, $i=1,2,3$. (b) A quasi-metallic zigzag CNT.
}
\label{fig:chord}
\end{figure}
Upon ``rolling" a graphene sheet to form a nanotube it is convenient to rotate the coordinate system such that the $x$-axis lies along the chiral vector defined as  $\boldsymbol{C}_h=n\mathbf{a}_{1}+m\mathbf{a}_{2}$, where  $\mathbf{a}_{1}=\boldsymbol{R}_1-\boldsymbol{R}_2$ and  $\mathbf{a}_{2}=\boldsymbol{R}_1-\boldsymbol{R}_3$ are the primitive lattice vectors of graphene and that the $y$-axis lies along the nanotube axis. The nearest neighbor vectors in the rotated frame can be written as $(R_{C_i},R_{T_i})$, with the components given by the expressions:
\begin{equation}
\begin{aligned}
R_{C_i}=&R_{x_i}\cos \phi\ +\ R_{y_i}\sin \phi,
\\
R_{T_i}=&-R_{x_i}\sin \phi\ +\ R_{y_i}\cos \phi,
\end{aligned}
\label{eq:triad_new}
\end{equation}
where $R_{x_i}$ and $R_{y_i}$ are the Cartesian components of the $i^{th}$ nearest neighbor vector, and $\cos\phi=\sqrt{3}\left(n+m\right)a/(2\left|\boldsymbol{C}_h\right|)$ and $\sin\phi=\left(n-m\right)a/(2\left|\boldsymbol{C}_h\right|)$. We shall denote the component of the wavevector which lies along the circumference (that which is quantized) and the component of the wavevector which lies along the nanotube axis (that which is free) as $k_{C}$ and $k_{T}$ respectively. Applying the periodic boundary condition to the circumferential wavevector yields $k_{C}=2\pi l / \left|\boldsymbol{C}_h\right|$, 
where $l$ is an integer and plays the role of the particles angular momentum. For a quasi-metallic tube we set $l=s\left(n-m\right)/3$, where $s=\pm1$, and in the absence of curvature the crossing of the conduction and valance bands occurs at $s 2\pi (n+m)/ (\sqrt{3} \left|\boldsymbol{C}_{h}\right|)$. It should be noted that we allow both positive and negative values of $n$ and $m$, whereas traditionally they are both chosen positive.

Let us now consider the role of curvature. In rolling a graphene sheet
to form a carbon nanotube one decreases the length of the nearest
neighbor vectors. This is because the new distance is given by the
chord between the two sites:
\begin{equation}
\sqrt{4r^{2}\sin^{2}\left(\frac{R_{C_i}}{2r}\right)+R_{T_i}^2},
\end{equation}
where $r$ is the radius of the nanotube given by $r=a\sqrt{n^{2}+m^{2}+nm}/2\pi$. To understand the effects of including curvature, one can imagine the nanotube as an unrolled graphene sheet  (see Fig.\,\ref{fig:chord}\,(a)), defined by the modified set of lattice vectors $\tilde{\mathbf{a}}_{1}=\widetilde{\boldsymbol{R}}_{1}-\widetilde{\boldsymbol{R}}_{2}$ and $\widetilde{\mathbf{a}}_{2}=\widetilde{\boldsymbol{R}}_{1}-\widetilde{\boldsymbol{R}}_{3}$, where $\widetilde{\boldsymbol{R}}_{i}$ are the modified nearest neighbor vectors, defined as:
\begin{equation}
\widetilde{\boldsymbol{R}}_{i}=
\left( 
2r\sin\left(\frac{R_{C_i}}{2r}\right),R_{T_i}
\right),
\label{eq:triad_new_sheet}
\end{equation}
and therefore the new effective chiral vector is defined as $\widetilde{\boldsymbol{C}}_h=n \widetilde{\mathbf{a}}_{1}+ m \widetilde{\mathbf{a}}_{2}$,
and the quantization wavevector changes from $k_{C}=2\pi l /\left|\boldsymbol{C}_{h}\right|$ to $k_{C}=2\pi l /\left|\widetilde{\boldsymbol{C}}_{h}\right|$.
In the nearest-neighbor tight binding approximation the influence of a magnetic field is accounted for by adding the number
$f=\Phi/\Phi_0$ (here $\Phi$ is the magnetic flux through the polygons cross section rather than the original circular cross-section and $\Phi_0=h/e$ is the magnetic flux quantum) to the angular momentum quantum number $l$.~\cite{ReichBook} For example, one can see from Fig.\,\ref{fig:chord}\,(b) that upon rolling a quasi-metallic zigzag nanotube, $\left|\boldsymbol{R}_{1}\right|$ remains unchanged whereas the magnitude of $\boldsymbol{R}_{2}$ and $\boldsymbol{R}_{3}$ are reduced in comparison to that of a planar graphene sheet. 

The effect of curvature is to break the symmetry between the nearest
neighbor vectors, therefore breaking the former equivalency of $t_{1}$,
$t_{2}$ and $t_{3}$. The probability of hopping between
sites is inversely proportional to the distance squared between hoping
sites,~\cite{HarrisonBook1989} i.e. $t_{i}\propto\left|\boldsymbol{R}_{i}\right|^{-2}$, therefore the modified matrix elements of hopping, $\tilde{t}_{i}$,
are related to the original elements, $t_{i}$, by the simple expression:
\begin{equation}
\frac{\tilde{t}_{i}}{t_{i}}=\left(\frac{\left|\boldsymbol{R}_{i}\right|}{\left|\widetilde{\boldsymbol{R}}_{i}\right|}\right)^{2}.
\nonumber
\end{equation}
In the presence of curvature and applied magnetic field the modified electron energy spectrum for a quasi-metallic nanotube is given by:
\begin{equation}
\xi=\tilde{s}\left|\sum_{i=1}^{3}\tilde{t_{i}}\exp\left(i \widetilde{\boldsymbol{k}} \cdot \widetilde{\boldsymbol{R}}_{i}
\right)\right|,
\label{eq:energy_curvature_full}
\end{equation}
where $\widetilde{\boldsymbol{k}}$ is the charge carrier's wavevector of the effective graphene sheet,  hence 
$\widetilde{\boldsymbol{k}} \cdot \widetilde{\boldsymbol{R}}_{i}=k_{C}\widetilde{R}_{x_i}+k_{T}\widetilde{R}_{y_i}$, where $k_{C}=2\pi \left[f+s\left(n-m\right)/3\right]/\left|\widetilde{\boldsymbol{C}}_{h}\right|$ and $\widetilde{R}_{x_i}$ and $\widetilde{R}_{y_i}$ are the Cartesian components of the effective graphene sheet given in Eq.~(\ref{eq:triad_new_sheet}).

\begin{figure}
\centering
\includegraphics[width=0.8\linewidth]{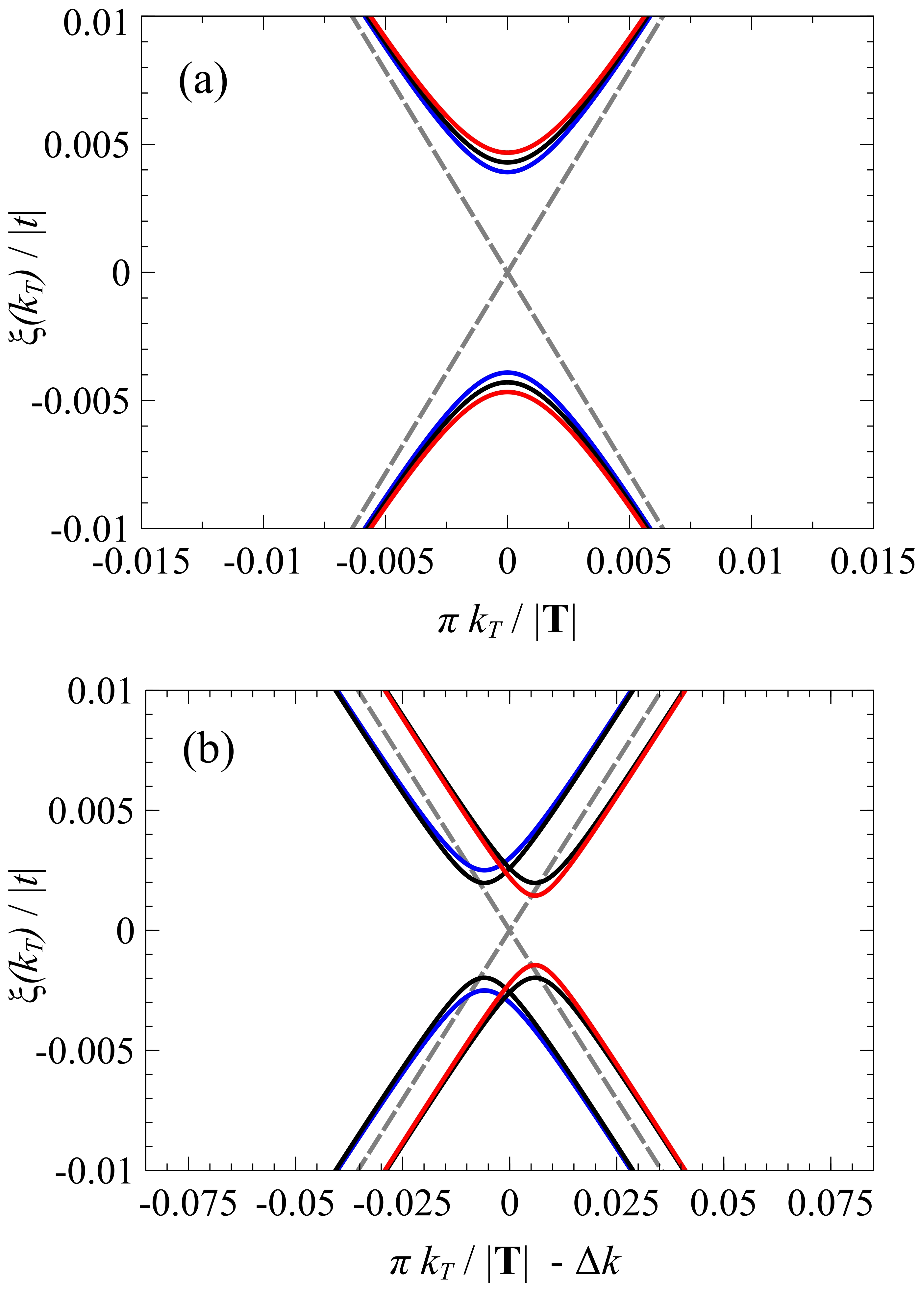}
\caption{
Detailed  view  of  the  gap  for (a)  a (12, -12) CNT and (b) a (15, 3) CNT, with (red and blue lines, corresponding  to  $s=+1$ and $s=-1$ respectively, 
for  the  case  of $B=5$  T)  and  without  (black line) an external magnetic field along the CNT axis.  The  dashed grey  line  corresponds  to  zero  field  
and no curvature. $\Delta k = s 2\pi (n+m)/ (\sqrt{3} \left|\boldsymbol{C}_{h}\right| ) $
}
\label{fig:Band_structure}
\end{figure}


Explicit expressions for the dispersion relations can be obtained for both zigzag quasi-metallic and armchair nanotubes. Let us first consider the case of zigzag quasi-metallic tubes defined by $(n,-n)$ where $n$ is a multiple of three. It should be noted that this is equivalent to a zigzag tube defined by $(n,0)$ with $l=s2n/3$. By symmetry, the hopping parameter $t_{1}$ does not change, i.e. $t_{1}=t$, while $t_{2}$ and $t_{3}$ are modified in the same way. In the presence of curvature and applied magnetic field the modified electron energy spectrum for a quasi-metallic zigzag CNT is given by:
\begin{equation}
\xi=\tilde{s}\left|t\right|\sqrt{\left(\frac{\xi_{g}}{2t}\right)^{2}+4\lambda_{\mathrm{z}}\sin^{2}\left(\frac{\sqrt{3}}{4}a k_T \right)},
\label{eq:energy_curvature_zigzag}
\end{equation}
where $\lambda_{\mathrm{z}}=-2\cos\left(\frac{\pi}{n}f \pm  \frac{2\pi}{3}\right)\tilde{t}_{2}/t$, $\xi_{g}=4\hbar v_{\mathrm{F}}\left|1-\lambda_{\mathrm{z}}\right|/ \left( 3a_{\mathrm{CC}} \right)$ and $v_{\mathrm{F}}=\sqrt{3}a\left|t\right|/\left(2\hbar\right)$ is the Fermi velocity of graphene. With the inclusion of curvature the spectrum is no longer linear near the crossing point (see Fig.~\ref{fig:Band_structure}) and a band gap of $\xi_{g}$ has appeared, whose size can be tuned by the strength of the applied magnetic field. For large radius tubes the band gap is given by $\xi_{g}\approx\hbar v_{\mathrm{F}}a_{\mathrm{CC}}\left|\frac{1}{16R^{2}}+s\frac{4}{\sqrt{3}a_{\mathrm{CC}}^{2}}\sin\left(\frac{\pi}{n}f\right)\right|$.
It should be noted that assuming a different power dependence of the transfer integral on the bond length results in the same dependence of the gap size with radius \cite{Kane1997,Kleiner2000} however, the magnitude of the gap varies by a geometric factor. In the absence of an applied magnetic field the energy spectrum is degenerate in $l$, this degeneracy is broken for any size magnetic field hence an applied magnetic field results in two separate bandgaps. For a $\left(12,-12\right)$ CNT in zero field this gap corresponds to $\approx 6.2$~THz and in the presence of a $5$~T field the two gaps are $\approx 6.8$~THz and $\approx 5.7$~THz which correspond to $s=1$ and $s=-1$ respectively. The low energy spectrum takes the form
\begin{equation}
\xi=\tilde{s}\sqrt{\left(\frac{\xi_{g}}{2}\right)^{2}+\lambda_{z} v_{\mathrm{F}}^{2}\hbar^{2}k_T^{2}}.
\label{eq:low_energy_curv}
\end{equation}
Eq.~(\ref{eq:low_energy_curv}) is similar in form to that of a one dimensional, massive, relativistic
Dirac fermion, and in the limit that $\tilde{t}_{2}\rightarrow t_{2}$
(i.e. neglecting the effects of curvature) Eq.~(\ref{eq:low_energy_curv})
restores the linear disperision of the simple zone-folding model of $\pi$-electron graphene spectrum.

For an armchair nanotube defined by $(n,n)$ the low-energy spectrum in a magnetic field becomes
\begin{equation}
\xi=\tilde{s} 
\sqrt{\left(\frac{\xi_{g}}{2}\right)^{2}+\lambda_{\mathrm{a}}v_{\mathrm{F}}^{2}\hbar^{2}\kappa^{2}},
\end{equation}
where $\kappa=k_{T}\mp2\arccos\left[\tilde{t}_{1}\cos\left(\frac{\pi}{n}f\right)/\left(2\tilde{t}_{2}\right)\right]/a$, the $\mp$ sign corresponds to the two different valleys, $\lambda_{\mathrm{a}}=\frac{4}{3}\left(\tilde{t}_{2}/t\right)^{2}-\frac{1}{3}\left(\tilde{t_{1}}/t\right)^{2}\cos^{2}\left(\frac{\pi}{n}f\right)$
and
\begin{equation}
\xi_{g}=\frac{4\hbar v_{\mathrm{F}}}{3a_{\mathrm{CC}}}\left|\frac{\tilde{t}_{1}}{t}\sin\left(\frac{\pi}{n}f\right)\right|.
\label{eq:gap_arm}
\end{equation}
Due to symmetry, curvature alone does not open up a gap.~\cite{Kane1997} However, curvature effects do result in the shifting of the crossing points, and an applied magnetic field also shifts the minimum and opens the gap equally for both valleys. It should be noted that in the limit that $\tilde{t}_{1}\rightarrow t$, Eq.~(\ref{eq:gap_arm}) restores the result obtained in Refs. \cite{Portnoi2008,PORTNOI2009,Hartmann2011}

For a chiral quasi-metallic tube, the effect of curvature is to open a band gap. Also their  minima deviate from the crossing points obtained in the absence of curvature effects. Much like zigzag quasi-metallic CNTs, when a magnetic field is applied along the nanotube axis, the band gaps for each valley are modified differently from one another (see Fig.~\ref{fig:Band_structure} (b)).

\subsection{Optical selection rules}
\begin{figure}
\centering
\includegraphics[width=0.8\linewidth]{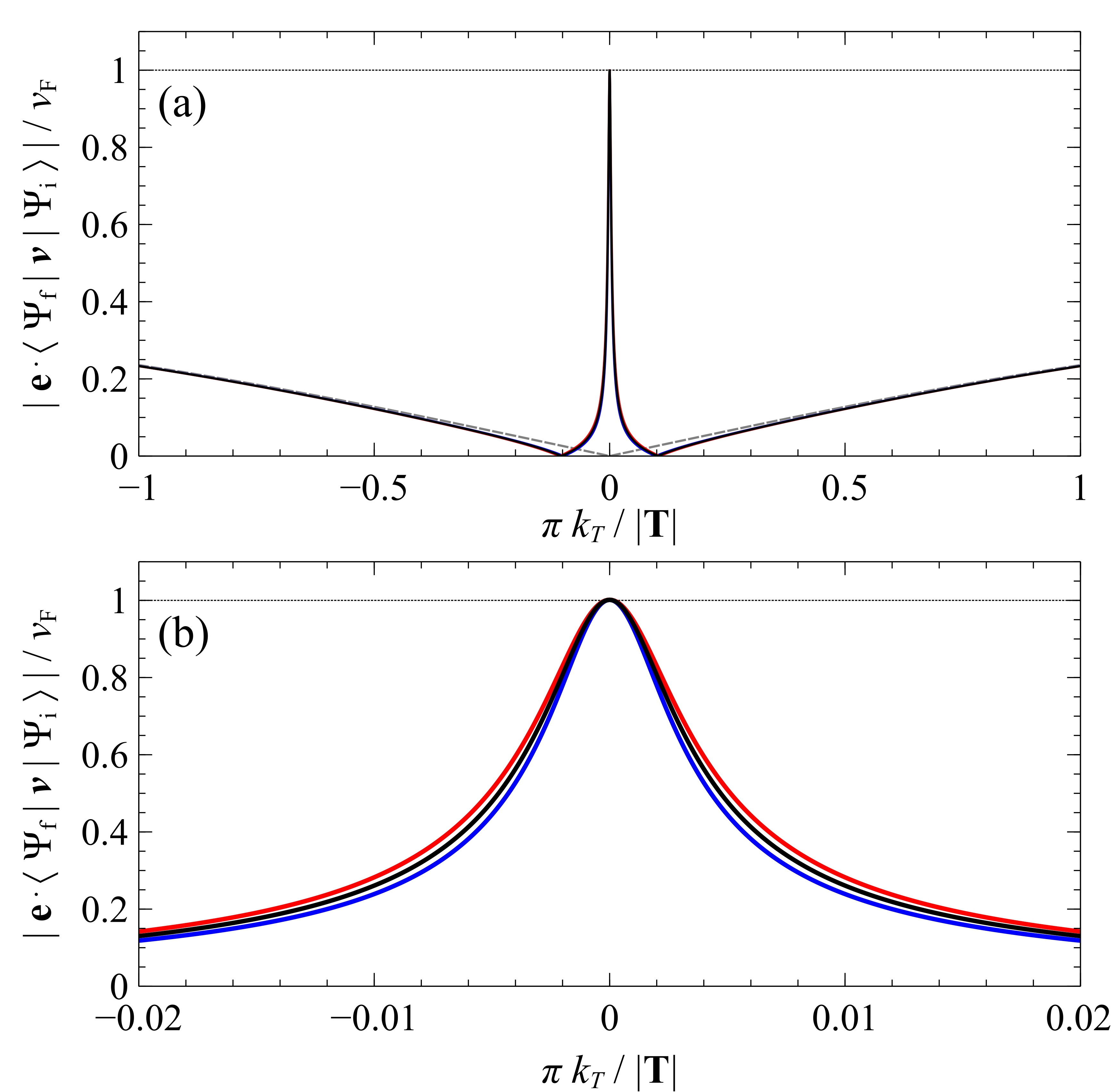}
\caption{
Dependence of the dipole matrix element for the transition between the top valence and lowest conduction subbands on the 1D wave vector $k_T$ for (a), across the Brillouin zone and (b), in the region close to the band gap edge, for a zigzag quasi-metallic tube defined by (12,-12) for $B=0$~T (black line) and $B=5$~T (red and blue lines, corresponding to $s=1$ and $s=-1$ respectively) applied along the CNT axis. The  dashed grey  line  corresponds  to  zero field and no curvature, while the dotted horizontal line is a visual aid highlighting the maximal value of the matrix element of velocity attained at the band gap edge.}
\label{fig:Matrix}
\end{figure}

In the dipole approximation, the spectral density of spontaneous emission, $I_{\nu}$, is
given by \cite{Kibis2007}
\begin{equation}
\begin{aligned}
I_{\nu}=&
\frac{8\pi e^{2}\nu}{3c^{3}}\sum_{i,f}
\left|\mathbf{e}\cdot\left\langle\psi_{f}\left|\hat{\bf{v}}\right|\psi_{i}\right\rangle \right|^{2}\\
&\times
f_{e}\left(k_{i}\right)f_{h}\left(k_{f}\right)
\delta\left(\xi_{f}-\xi_{i}-h\nu\right),
\end{aligned}
\label{eq:Emission}
\end{equation}
where $\psi_{i}$ and $\psi_{f}$ are the eigenfunctions of the electrons in the initial and final states, $\xi_{i}$ and $\xi_{f}$ are their associated energies, and $k_{i}$ and $k_{f}$ are their associated wave vectors, $f_{e}$ and $f_{h}$ are the distribution functions of electrons and holes, $\hat{\bf{v}}$ is the velocity operator, $\mathbf{e}$ is the polarization of the excitation which we take to be propagating along the nanotube axis and $\nu$ is the frequency of the excitation. Using the velocity operator in commutator form: $\hat{\bf{v}}=\frac{i}{\hbar}\left[\mathcal{\hat{H}} ,\,\boldsymbol{r} \right]$, where $\mathcal{\hat{H}}$ is the tight-binding Hamiltonian of the modified graphene sheet~\cite{SemenoffPRL1984} described by the vectors $\widetilde{\boldsymbol{R}}_i$,
allows the the matrix element
$\mathbf{e}\cdot\left\langle \psi_{f}\left|\hat{\bf{v}}\right|\psi_{i}\right\rangle$
to be written as\cite{BookHartmann2011}
\begin{equation}
\mathbf{e}\cdot\Re
\left[
\frac{\tilde{f}_{k}^{\star}}
{
\hbar\left|\tilde{f}_{k}\right|
}
\sum_{i=1}^{3}\tilde{t}_{i}e^{i\boldsymbol{k}\cdot\widetilde{\boldsymbol{R}}_i}\widetilde{\boldsymbol{R}}_i
\right].
\label{eq:Mat_element}
\end{equation}
The same result can also be obtained within the gradient approximation.~\cite{Saroka2018MomentumAlign} Eq.~\ref{eq:Emission} and Eq.~\ref{eq:Mat_element} are sufficient to generate the spectral density of spontaneous emission for all quasi metallic tubes, with and without an applied magnetic field. For the case of zig-zag quasi-metallic CNTs, Eq.~\ref{eq:Mat_element} admits simple analytic expressions, in the presence of curvature and an applied magnetic field:
\begin{equation}
\frac{a_{\mathrm{CC}}}{8}\omega_{if}+\frac{2v_{\mathrm{F}}^{2}}{3a_{\mathrm{CC}}\omega_{if}}\left(1-\lambda_{\mathrm{z}}^{2}\right)
\label{eq:Curvature_element}
\end{equation}
Which at the band gap edge becomes $\approx v_{\mathrm{F}}$ for experimentally attainable magnetic fields and typical nanotube diameters, and in the absence of an applied magnetic field is zero as $\tilde{t}_{2}\rightarrow t$. Curvature not only opens the gap in the quasi-metallic zigzag CNT spectrum, but
also allows dipole optical transitions at the bandedge between the highest
valence subband and the lowest conduction subband. Indeed, in the THz regime the results obtained for zig-zag quasi metallic tubes, hold true across the entire class of quasi-metallic CNTs, differing only by geometrical factors. In Fig.~\ref{fig:Matrix} we show how the matrix element of the dipole optical transitions polarized along the CNT axis are modified in the presence of a magnetic field.

As we discussed above, for an $(n,n)$ armchair nanotube curvature alone is insufficient to open the gap. However, a longitudinal magnetic field applied along the nanotube axis not only opens a bandgap but gives rise to strong optical transitions at the band edge. The velocity operator in this instance is given by the simple analytic expression
\begin{equation}
\frac{8v_{\mathrm{F}}^{2}}{3\sqrt{3}a_{\mathrm{CC}}\omega_{if}}\left(\frac{\tilde{t}_{1}}{t}\right)^{2}\sin\left(\frac{\pi}{n}f\right)\sqrt{\left(\frac{\tilde{t}_{2}}{\tilde{t}_{1}}\right)^{2}-\frac{1}{4}\cos^{2}\left(\frac{\pi}{n}f\right)}
,
\label{eq:mat_arm}
\end{equation}
and at the band edge Eq.~\ref{eq:mat_arm} is $\approx v_{\mathrm{F}}$ for experimentally attainable magnetic fields with a typical nanotube diameter.

\begin{figure}
\centering
\includegraphics[width=0.8\linewidth]{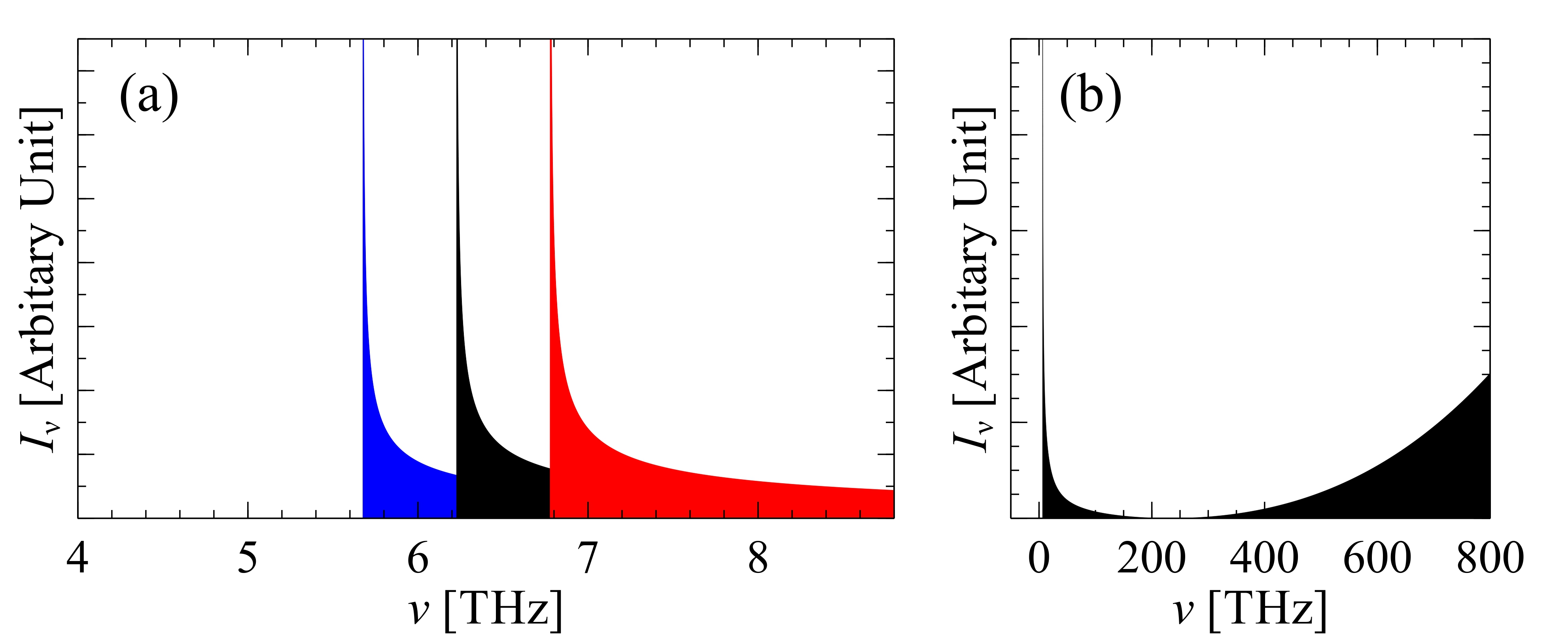}
\caption{
The  calculated  photon emission  spectra  for  a  (12,-12)  CNT  in  zero  magnetic field  (black curve)  and  a  magnetic field  of  $B=5$ T, where the red and blue curves correspond to $s=1$ and $s=-1$ respectively, for (a) frequencies of the order of the band gap and (b) for a broad range of frequencies. 
}
\label{fig:vanhove}
\end{figure}

The spectral density of spontaneous emission for a zigzag quasi-metallic CNT taking into account curvature effects and applied magnetic field is obtained by substituting Eq.~(\ref{eq:Curvature_element}) into Eq.~(\ref{eq:Emission}) then performing the necessary summation. In the THz regime one obtains the expression:
\begin{equation}
I_{\nu}=Lf_{e}\left(k_{i}\right)f_{h}\left(k_{f}\right)\frac{\pi^{3}e^{2}a_{\mathrm{CC}}^{2}\left[12t^{2}\left(1-\lambda_{\mathrm{z}}^{2}\right)+h^{2}\nu^{2}\right]^{2}}{3c^{3}h^{4}v_{\mathrm{F}}\sqrt{\lambda_{\mathrm{z}}\left(h^{2}\nu^{2}-\xi_{g}^{2}\right)}} \, ,
\label{eq:curvature_spectral}
\end{equation}
where $L$ is the tube length. In the absence of an applied magnetic field, the electronic (hole) energy spectrum near the bottom (top) of the conduction (valence) band is no longer linear due to curvature effects, and the van Hove singularity in the joint density of states leads to a very sharp absorption maximum near the band edge and correspondingly to a very high sensitivity of the photocurrent to photon frequency, see Fig.~\ref{fig:vanhove}. In the presence of an applied magnetic field, the absorption peak associated with curvature-induced transitions is split, the two absorption maxima corresponding to $s=1$ and $s=-1$ (see Fig.~\ref{fig:vanhove}), respectively. These results hold true across the whole class of quasi-metallic tubes. 

With the knowledge of the curvature-induced or magnetic-field-induced gap and strength of the associated transitions, we must review the earlier proposed scheme for THz radiation generation by hot carrier recombination in narrow-gap CNTs.~\cite{Kibis2007} In fact the injection scheme does not require high voltage since the overlooked band-edge-transitions are strongly allowed. One must only overcome the Zener breakdown voltage which is used for the electric injection of carriers into the nanotube conduction band. 

\begin{figure}
	\includegraphics*[width=0.75\linewidth]{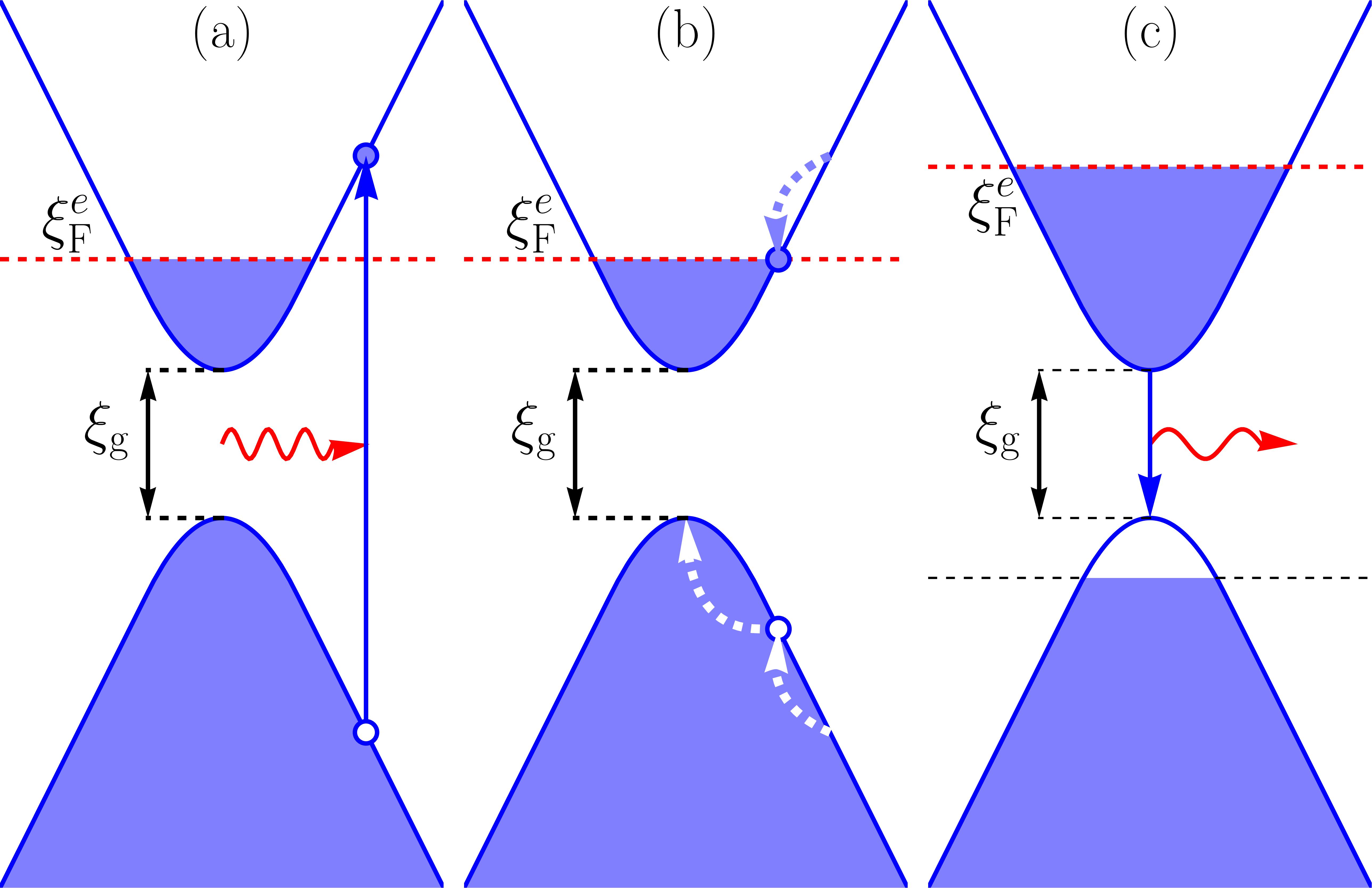}
	\caption{%
		A schematic illustration of (a) the high frequency optical excitation (b) non-radiative electron relaxation due to the electron-phonon scattering and (c) the population inversion in an $n$-doped narrow gap CNT or GNR. The red-dashed line depicts the non-equilibrium quasi Fermi level, $\xi_\mathrm{F}^e$, in the conduction band.}
	\label{fig:Excitation_scheme}
\end{figure}

Another, fully-optical scheme for the observation of strong band-edge transitions in narrow-gap CNTs can be proposed.
We must stress that we do not question the origin of the broad THz absorption peak observed in spectroscopy experiments being of plasmonic nature. Indeed any spurious doping results in the suppression of curvature induced interband absorption because of the absence of empty states near the edge of the nanotube conduction band (Pauli blocking). 
Therefore, we suggest that the band-edge transitions should be observed in THz emission experiments rather than absorption ones. To detect the reported feature one can create a population inversion, for example, by optical pumping (see Fig.~\ref{fig:Excitation_scheme}). It can be seen from Eq.~(\ref{eq:curvature_spectral}) that at optical frequencies, the transition probability is proportional to the cube of frequency. Therefore using a broad range of optical frequencies will lead to the effective promotion of many electrons into the conduction band, and the creation of holes in the valance band. The photoexcited electrons and holes quickly thermalize with the lattice due phonon scattering ($\tau_{ph} \sim 3$~ps).~\cite{Park2004} As a result of this process the holes move up to the top of the valence band, creating a population inversion, and the excited electrons join the Fermi sea contributing to the increase of the non-equilibrium quasi Fermi level in the conduction band. The emission of photons of the band-gap frequency will occurs with an extremely high probability, since the optical matrix element is maximal at the band gap edge and the density of states diverges. Therefore, this effect can be used for the generation of a very narrow emission line having the peak frequency tunable by the applied magnetic field. The emission output can be maximized by putting an array of narrow-gap CNTs into a microcavity similar to what has been done for semiconducting CNTs.~\cite{He2018,Gao2018} THz mirrors with low losses should be carefully designed~\cite{Headland2017} to achieve gain in this case. The analysis of losses in such a system will be reported elsewhere. It should be noted that the absorption by free carriers in the sample can be minimized by the proper choice of structure length, since the plasmonic resonance is a geometrical one and depends strongly on the structures longitudinal size.~\cite{Slepyan2010,Shuba2012}
Since semiconducting tubes are transparent to THz radiation, there is no need to separate the semiconducting from the quasi-metallic CNTs. However, ideally samples should be enriched with narrow-gap nanotubes. 

\section{Graphene nanoribbons}\label{sec:GNRs}

The band structure of AGNRs can be obtained from that of graphene by a technique similar to that used with CNTs. The periodic boundary condition applied to the tube, $\boldsymbol{k}_{T}\cdot \boldsymbol{C}_h =  2 \pi l$, is replaced with the so-called ``hard wall" or ``fixed ends" boundary condition, $\boldsymbol{k}_{T}\cdot \boldsymbol{L} =  \pi l$, where $L$ is the ribbon's width and $\boldsymbol{k}_{T}$ is the electron's transverse momentum. It should be noted that these two types of boundary conditions match if $L= C_h / 2$, this occurs for example for AGNR$(N)$ and zigzag CNT$(N+1,0)$.~\cite{White2007,SAROKA2018JSCS} For these specifically chosen structures the electronic properties are almost identical. At low energies, the band spectra of these tubes are almost an exact replica of that of the ribbons, the only difference is that the tubes bands are double degenerate, whereas the ribbons bands are not. However, at higher energies the band structures deviate from one another, and the spectrum of tubes contains some higher energy bands which are absent in the ribbons. The edge effect in armchair ribbons can be incorporated into the tight-binding model as corrections to the hopping integrals at the ribbon edges.~\cite{Zheng2007} 

\begin{figure}
	\includegraphics*[width=0.8\linewidth]{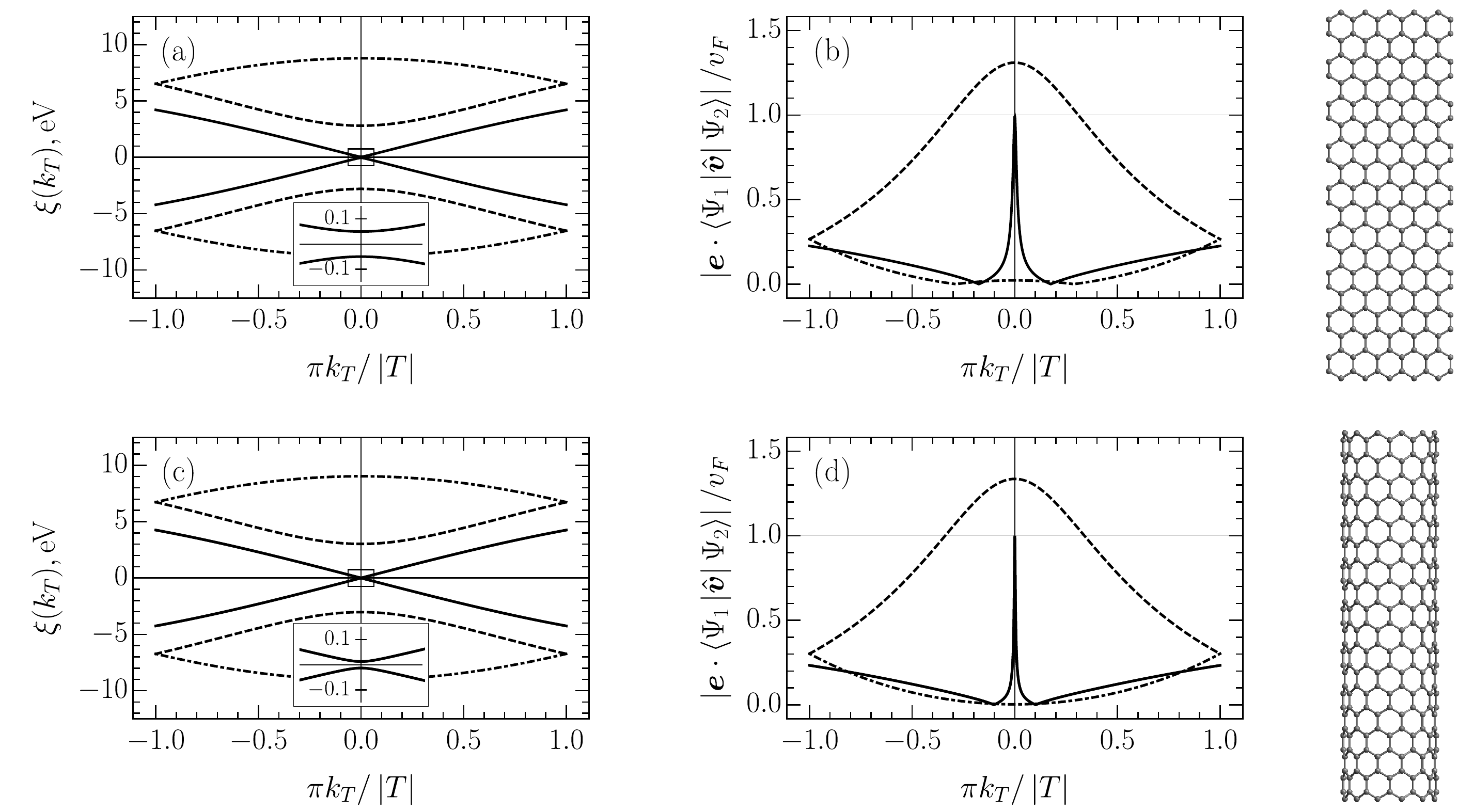}
	\caption{
(a),(c) selected bands from the electronic band structure and (b),(d) the velocity operator matrix elements normalized by $v_F$ of an AGNR$(11)$ and zigzag CNT$(12,-12)$, respectively.  Transitions between the closest valence and conduction subbands (thick black), the lowest and highest subbands (dashed dotted, light gray), and for the subbands, for which matrix element of velocity attains the maximum possible value (dashed, gray) are shown. The insets (a) and (c) show the zoomed in region close to the Dirac point where the band gap is present. On the right side the atomic structures are shown. In both cases the hopping integral, $t=3$~eV and the edge correction for the ribbon is $0.05 t$.
	}
	\label{fig:AGNRvsZCNT}
\end{figure}

In Fig.~\ref{fig:AGNRvsZCNT} (a), (c) we show selected bands from the electronic band structure of a AGNR and a zigzag CNT for the case of $N=11$, taking into account the edge effect in the ribbon and the curvature effect in the tube. It can be seen from the figure that the described equivalence of the low energy band spectra is held throughout the whole band structure. This equivalence extends to optical transitions selection rules. Our calculations, presented in Fig.~\ref{fig:AGNRvsZCNT} (b), show that the edge effect for quasimetallic armchair GNRs results in a peak similar to that in Fig.~\ref{fig:AGNRvsZCNT} (d). As was the case with narrow-gap CNTs the peak has the same characteristic height, equal to the $v_F$ the bandgap edge. This coupled with the presence of the Van Hove singularity gives rise to a large interband transition probability rate. Thus like narrow-gap CNTs, AGNR are promising candidates for the active element in THz emitters. Their emission frequency can be tuned by applying an in-plane electric field.~\cite{Chang2006,Saroka2015} It should be noted that synthesis techniques for such structures are developing at a fast pace, for example AGNRs of the metallic family can already be produced with atomically smooth edges.~\cite{Zhang2015} Therefore, there is much promise that ideal samples are on the horizon.

It is worth emphasizing that the equivalence between the optical properties of tubes and ribbons reported herein is not trivial. Although the curvature effect in tubes and the edge effect in ribbons both represent an intrinsic strain, the former is a homogeneously distributed over the tube surface, while the latter is localized at the ribbon edges. Finally, it should be noted that the band gap of metallic AGNRs are also influenced by third order nearest neighbours terms (3NN).~\cite{White2007} Within the framework of the analytical model proposed  Gunlucke,~\cite{Gunlycke2008} it can be shown that the inclusion of the 3NN hoping integrals in the consideration of metallic AGNRs results in a peak in the transition probability similar to the 1NN model. However, the Gunlycke model reproduces only transition probabilities between highest valence and the lowest conduction subbands of the Zheng model~\cite{Zheng2007} giving qualitatively different transition probability rates of higher energy transitions. The two pictures are yet to be reconciled.

\section{Excitonic Effects}\label{sec:Exciton}

The results presented in Sec.~\ref{sec:CNTs} and \ref{sec:GNRs} are based on a single electron picture. In this section, will shall discuss the role of excitonic effects in narrow-gap CNTs and GNRs. Many-body  (excitonic)  effects, are known to dominate  the  optical  properties  of
semiconducting CNTs~\cite{Ando97JPSP,Wang2005a,Maultzsch2005,Shaver2007} and result in extremely low optical quantum yields. The suppression of photoluminescence is due to the presence of dark excitonic states, these non-radiative states have a significantly lower energy than the radiative bright excitonic states.~\cite{spataru2004excitonic,perebeinos2004scaling,perebeinos2005radiative,scholes2007low} Therefore, bright excitons relax towards the dark state, and consequently non-radiative decay dominates over the radiative.~\cite{Avouris07NatTech,Avouris08NatPhoton} Several methods have been proposed to enhance the luminescence efficiency in semi-conducting tube,~\cite{kilina2012brightening,harutyunyan2009defect} including the use of a microcavity \cite{miura2014ultralow,luo2017purcell,jeantet2017exploiting} and magnetic brightening.~\cite{Shaver2007, shaver2007magnetic, Srivastava2008} In quasimetallic CNTs and ANGRs the exciton binding energy has been shown to never exceed the bandgap for both long-range~\cite{Hartmann2011,Portnoi2013IEEE} and short-range interaction potentials.~\cite{Portnoi2013IEEE,Hartmann2017} Therefore, unlike semi-conducting tubes, the electron-hole pairs should be fully ionized at room temperature. Hence, the aforementioned undesirable effects due to dark excitons, should not dominate the optical processes in narrow-gap nanotubes. However, carrier interaction is still important. In quasi-one-dimensional semiconductors, the introduction of coulomb interaction between carriers results in the absence of the singularity associated with band edge transitions in the absorption spectrum.~\cite{haug2009quantum}
Initial studies of quasi-one dimensional narrow-gap structures show that both long-range and short-range interaction models~\cite{Portnoi2013IEEE} also result in the suppression of the van Hove singularity.

In THz absorption experiments, the combination of the van Hove singularity suppression, Pauli blocking caused by spurious doping, and a fast reduction in the interband transition matrix element away from the band gap edge results in the absorption peak being purely plasmonic in nature. However, the situation becomes different for the optically-induced population inversion scheme discussed at the end of Section \ref{sec:CNTs} (see Fig.\,\ref{fig:Excitation_scheme}). Since the screening is much weaker in one-dimensional systems compared to  bulk  or even two-dimensional materials, the excitonic peak should persist in the presence of free carriers.
A possible consequence could be two close THz emission peaks (which will be arguably difficult to resolve given the current state of THz spectroscopy). The very narrow peak at a lower energy will be produced by an optically active excitonic state below the band gap~\cite{Hartmann2017}; whereas, the higher-energy broader peak should occur slightly above the band gap edge - it results from the combination of the band-edge van Hove singularity suppression and the decay of the matrix element with increasing photon energy.
The in-depth study of both the van Hove singularity suppression and excitonic transitions in ultra-relativistic quasi-one-dimensional systems remains a subject of current research.


\section{Conclusions}
In the absence of curvature or edge effects, optical interband transitions near the crossing points of the valence and conduction bands of metallic CNTs and AGNRs are vanishing with reducing frequency. The effects resulting in the opening of a gap, which is typically in the THz range, also lead to a drastic change of the wavefunctions near the band gap edges which in turn allows optical transitions. These transitions are very strong at low frequencies and their matrix elements are several orders of magnitude larger then those previously calculated in a model which neglects curvature.~\cite{Kibis2007}  Furthermore, the frequency peaks in the spectral density of emission in narrow-gap CNTs can be tuned by the application of a magnetic field directed along the nanotube axis. For both quasi-metallic CNTs and GNRs the edge optical transition frequency can be modified by an in-plane electric field via the Franz-Keldysh effect: this field can be induced, e.g., by a split back-gate below the substrate underneath the CNT or GNR array. 
Appropriately arranged arrays of CNTs or GNRs should be considered as promising candidates for active elements of amplifiers and generators of coherent THz radiation. In nanotube arrays, all quasi-metallic CNTs with the same chirality will emit in a similar fashion, whereas semiconducting and armchair nanotubes (in the absence of an applied magnetic field) will be optically inactive in the THz and mid-infrared range. In addition, the discussed effects provide a spectroscopic tool allowing to differentiate between quasi-metallic and true metallic quasi-one-dimensional carbon nanostructures.

\begin{acknowledgments}
This work was supported by the EU FP7 ITN NOTEDEV (FP7-607521); EU H2020 RISE project CoExAN (H2020-644076); FP7 IRSES projects CANTOR (FP7-612285), QOCaN (FP7-316432), and InterNoM (FP7-612624). RRH acknowledges financial support from URCO (17 F U 2TAY16-2TAY17). The work of MEP was supported by the Government of the Russian Federation through the ITMO Fellowship and Professorship Program. MEP is also grateful for hospitality at the International Center for Polaritonics of Westlake University.  
\end{acknowledgments}

%


\end{document}